\documentclass[aps, twocolumn,showpacs,preprintnumbers,amsmath,amssymb]{revtex4}

\usepackage{graphicx}
\usepackage{dcolumn}
\usepackage{bm}
\usepackage{epstopdf}
\usepackage{MnSymbol}

\begin{document}

\preprint{APS/123-QED}

\title{Quantum transport in mesoscopic $^3$He films: experimental study of the interference of bulk and boundary scattering}

\author{P. Sharma$^{\dagger}$, A. C\'{o}rcoles$^{\dagger,\star}$,  R.G. Bennett$^{\dagger,\spadesuit}$,
J.M. Parpia$^{\spadesuit}$, B. Cowan$^{\dagger}$, A. Casey $^{\dagger}$, J. Saunders$^{\dagger}$ }

\affiliation{$\dagger$Department of Physics, Royal Holloway University of London, Egham, TW20 0EX, Surrey, UK}
\affiliation{$\star$Present address, IBM Watson Research Center, Yorktown Heights, NY 10598, USA}
\affiliation{$\spadesuit$Department of Physics, Cornell University, Ithaca, NY, 14853 USA}

\date{\today}

\begin{abstract}
We discuss the mass transport of a degenerate Fermi liquid $^3$He film over a rough surface, and the film momentum relaxation time, in the framework of theoretical predictions. In the mesoscopic r\'egime, the anomalous temperature dependence of the relaxation time is explained in terms of the interference between elastic boundary scattering and inelastic quasiparticle-quasiparticle scattering within the film. We exploit a quasiclassical treatment of quantum size effects in the film in which the surface roughness, whose power spectrum is experimentally determined, is mapped into an effective disorder potential within a film of uniform thickness. Confirmation is provided by the introduction of elastic scattering centres within the film. We model further studies on $^3$He confined in nanofluidic sample chambers with lithographically defined surface roughness. The improved understanding of surface roughness scattering may impact on enhancing the conductivity in thin metallic films.
\end{abstract}
\maketitle

The factors determining electrical transport in thin metallic films are a long-standing problem of both fundamental interest and technological importance~\cite{Josell}. Thin metal nanowires provide the interconnects in electronic nanodevices, and the improvement of their performance should lead to reduced power loss, delays etc. The resistivity of films of ultra-high purity and quality will inevitably be limited by surface scattering. The work reported here uses thin $^3$He films as a simple model system to test the available theories of quantum transport in degenerate Fermi films, and contributes to establishing a first principles understanding of surface scattering.

Since the first theoretical discussions of size effects in thin films~\cite{Fuchs,Sondheimer}, in which surface scattering is treated as a phenomenological boundary condition imposed on the bulk transport equation, there have been significant theoretical developments which fully consider the importance of quantum size effects \cite{Tesanovic86,TrivediAshcroft88}. In a uniform film of thickness $L$, motion transverse to the film is quantized in particle-in-a-box states with $p_z  = \pi \hbar j/L$ resulting in a set of two-dimensional minibands with energy $\varepsilon  = \frac{1}{{2m^* }}[p_z^2  + q^2 ]$, in the simplest case of free fermions. In the presence of surface roughness the film thickness is non-uniform and $L \to L - \xi (x,y)$. This system can be mapped back into a film of uniform thickness by a canonical transformation, thereby introducing an effective disorder potential within the film, fully determined by $\xi (x,y)$. The influence of surface scattering is then calculated by incorporating this disorder potential into the film transport equation~\cite{MeyerovichStepaniants,MeyerovichPonomarev, MeyerovichChatterjee}. It is remarkable that this procedure allows an essentially first-principles calculation of the mesoscopic film conductivity (in the limit that elastic scattering processes within the film are negligible) in terms of the surface roughness power spectrum, which is independently experimentally measurable by scanning probe techniques.

These theoretical models predict that Matthiessen's rule may readily be violated in thin films. Experimental tests of this prediction in thin metallic films have focussed on studies of the resistivity of such films as a function of thickness \cite{Munoz,Feldman}. These studies are complicated by the presence of grain boundary scattering within the film~\cite{Sun}. However a further clear prediction of the quantum transport calculation is that, under the right conditions discussed below, there is a classical `interference' between elastic boundary scattering and inelastic quasiparticle-quasiparticle scattering, leading to anomalous temperature-dependence of the transport relaxation time and breakdown of Matthiessen's rule. It is the experimental observation and identification of this effect for the first time that is a significant result of our work.

The present study focuses on transport in thin films of $^3$He over rough surfaces, and flow through nanofluidic cavities. These provide model systems to investigate the interplay of surface scattering and bulk inelastic scattering, in the absence of grain boundary and imputity scattering. At temperatures below 100~mK this Fermi liquid is in the fully quantum degenerate regime ($T \ll T_\mathrm F$), and the bulk inelastic scattering time $\tau_\mathrm b$ arises from quasiparticle-quasiparticle scattering with $\tau_\mathrm b  \propto T^{ - 2}$ and mean free path $\lambda_\mathrm{in}  = 65/T^2 (\mu {\rm{m}}\,{\rm{mK}}^{\rm{2}} )$.

The experiments~\cite{Casey1} found a transport relaxation time that was unexpectedly long, manifesting surface scattering far weaker than we had anticipated, and exhibited an anomalous temperature dependence, $\tau _{{\rm{tr}}}  \propto T^{ - 1}$. In this paper we show that these results may be understood in terms of the quasi-classical treatment of film transport, referred to earlier. The anomalous temperature dependence provides direct confirmation of the predicted interplay between bulk and surface scattering. The `input' for our numerical calculations is the experimentally determined surface roughness power spectrum. Further experiments reported here confirm that adding random elastic scattering centres, by a surface decoration technique, significantly shortens the observed film-substrate relaxation time, as expected.

Momentum transfer between a film of fluid $^3$He and a surface can be studied at finite frequency using the torsional oscillator technique. The film couples to the transversely oscillating surface, contributing to the effective moment of inertia and dissipation of the oscillator, detected through changes in its frequency and quality factor. The film thickness is much smaller than the liquid viscous penetration depth. Thus the film behaves as a rigid object, coupled to the motion of the oscillator. The film--surface coupling is characterized by the momentum relaxation time of the film $\tau_\mathrm{osc} $, which is directly inferred \cite{Casey1} from the ratio of the film contribution to the dissipation of the oscillator and to its frequency shift $\omega \tau _{{\rm{osc}}}  = \Delta (Q^{ - 1} )/(\Delta f/f_0 )$.

\begin{figure}
    \includegraphics [width=8.1cm]{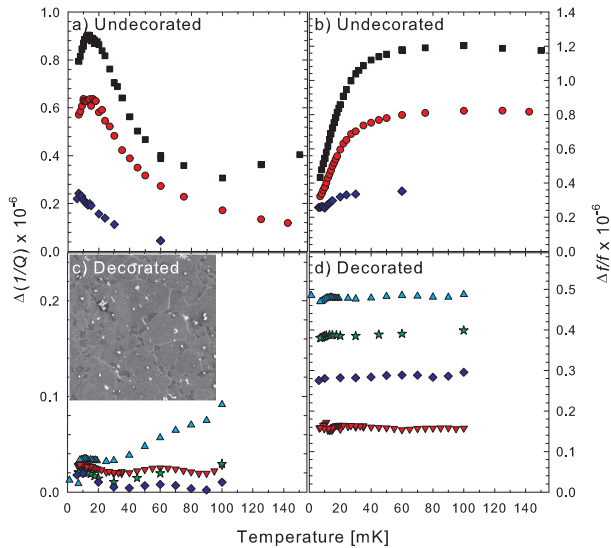}
    \caption{(color online)The frequency shift (right) and dissipation (left) from ``decorated" (bottom) and undecorated (top) silver surfaces. The nominal $^3$He film thicknesses are 350~($\filledsquare$), 240~($\bullet$), 143~($\filledmedtriangleup$), 112~($\filledstar$), 100~($\filleddiamond$) and 44~($\filledmedtriangledown$)~nm. The undecorated surfaces show the progressive decoupling of the inertia accompanied by a dissipative peak at a thickness dependent temperature. The decorated films show no such mass decoupling. Torsional oscillator frequencies are 2.84 kHz (undecorated), 2.57 kHz (decorated). Inset c) shows an optical micrograph ($100~\mu$m$\times 100~\mu$m)of the decorated silver surface.}
    \label{figure1}
    \end{figure}
In our prior experiment three films of $^3$He of nominal thickness 100, 240 and 350~nm were deposited on polished silver surfaces~\cite{Casey1}. We observed that the film decoupled from the motion of the oscillator; the key results are reproduced in Fig. 1 a), b). This was surprising given the atomically rough nature of the polished silver surface on which the $^3$He resided. Locally specular scattering of $^3$He quasiparticles from a rough surface, would give effectively diffuse scattering with high momentum transfer. The typical quasiparticle time-scale for film traversal is $\tau _{{\rm{film}}}  = L/v_\mathrm F$, giving 5~ns for a 300~nm film, or $\omega \tau _{{\rm{film}}}  \sim 10^{ - 4}$, suggesting strong and temperature independent film--substrate coupling. However the experimental finding is that $\tau _{{\rm{osc}}}  \gg \tau _{{\rm{film}}}$, at all temperatures investigated, increasing with decreasing temperature as $T^{ - 1}$ and passing through $\omega \tau _{{\rm{osc}}}  \sim 1$, yielding a characteristic dissipation maximum, see \cite{Casey1} for further details.
    As a control experiment we have repeated this earlier study with a new oscillator, which uses similarly-prepared polished surfaces decorated with discrete scattering centres, consisting of 700~nm diameter silver particles diffusion bonded/sintered to the silver surface. These are large defects, comparable to the film thickness. A defect correlation length of 7.85~$\mu $m was obtained from the analysis of optical micrographs, giving a temperature-independent elastic scattering time with $\omega \tau _{{\rm{osc}}}  \sim 2\times 10^{ - 3}$. As expected this surface treatment results in full coupling between the helium film and the oscillator. As shown in Fig. 1 c), d) no change in frequency shift is observed, as well as a smaller film contribution to the dissipation.

We now discuss how the film slip observed on undecorated mechanically-polished silver surfaces can be understood quantitatively in terms of the prediction of the surface scattering models, given the experimentally determined surface roughness power spectrum.

\begin{figure}[!h]
    \includegraphics [angle=0,width=8.1cm]{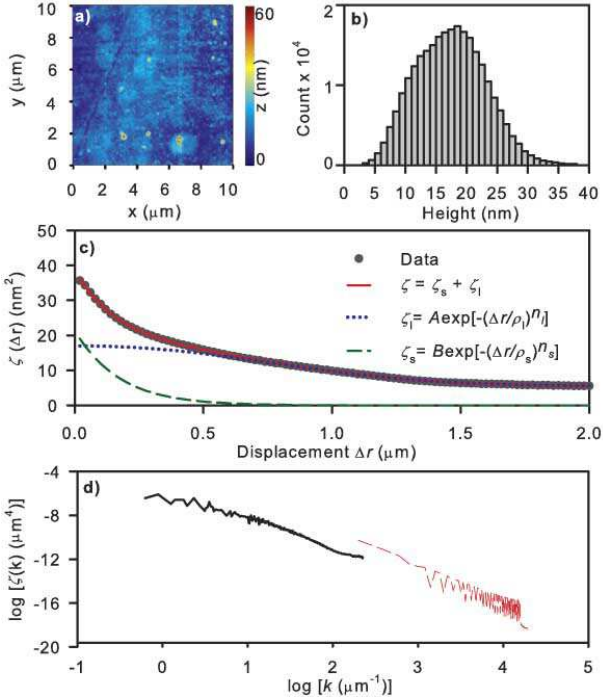}
    \caption{(color online)(a)Topographical AFM image of the undecorated silver surface. (b) Histogram of the height profile, standard deviation $\sigma=6$ nm. (c) The direct computation of the height-height correlations of the surface roughness, fit to a double stretched exponential, shows short- and long-range correlation lengths $\rho_{\rm{s}}=172$ and $\rho_{\rm{l}}=1005$~nm. In~(d) we show the power spectrum of the correlations measured for the undecorated silver surface (solid line) as well as for a lattice of square posts (dashed), with $h=8~ w=20~  p=40~ nm$, see text.}
    \label{figure2}
\end{figure}

The AFM scans of the polished silver surface reveal a nearly featureless surface (Fig.~\ref{figure2}a). The distribution of surface heights ($z$) is slightly skew, with a root-mean-square deviation, $l$, of 6~nm (Fig.~\ref{figure2}b).  The height--height autocorrelation function, $\zeta(\Delta x,\Delta y)$, is calculated as detailed in \cite{dimov}. The result, with $\Delta r = \sqrt{(\Delta x)^2 + (\Delta y)^2}$ to account for surface statistical isotropy, is shown in Fig. \ref{figure2}c. The autocorrelation function can be best fit by stretched exponentials over two different length scales. We found the short- and long-range correlation lengths, $\rho_{\rm{s}}$ and $\rho_{\rm{l}}$, to be 172 and 1005 nm, respectively. The surface roughness power `spectrum' $\zeta(\textbf{q})$ is the Fourier transform of the autocorrelation function $\zeta(\mathbf r)$.

We incorporate the experimental surface roughness power spectrum directly into the framework of the theory of Meyerovich~\cite{MeyerovichStepaniants} to numerically calculate the relevant transport time. The total number of minibands in the film is given by $j^*  = k_\mathrm F L/\pi$, where the Fermi wavevector $k_\mathrm F  = 7.85~{\rm{ nm}}^{{\rm{ - 1}}}$. Hence the number of mini-bands is large, and we are in the quasi-classical limit. Nevertheless quantum size effects manifest themselves through the effective disorder potential determined by $\xi (x,y)$. The quasi-classical transport equation is formulated for the effective uniform film, after the appropriate coordinate transformation, in terms of collision operators for both bulk inelastic scattering and surface scattering via the effective disorder potential. The expression for the effective relaxation rate of a film, $\tau_{\rm{eff}}^{-1}$, the imaginary part of the self-energy, is written~\cite{MeyerovichStepaniants} in terms of a boundary-induced transition probability $W(\bf{p},\bf{p'})$ for quasiparticles scattering off the effective disorder potential. This is given in terms of the surface roughness power spectrum $\zeta(\textbf{q})$.
\begin{equation}
\label{MeyerovichFormulaTauEff}
\frac{1}{\tau_{\rm{eff}}\!(\textbf{p})}\! =\! \frac {1}{\tau_{\rm{b}}}\! [\!1\!+L\!\!\!\int\nolimits\!\!\!\!\!\frac{d\textbf{p}'}{(2\pi \hbar)^3} \frac {W(\textbf{p},\textbf{p}')}{(\varepsilon(\textbf{p}')\! -\! E_\mathrm F )\!^2 \! /  (2 \pi \hbar ^2)\! +\! 1/4\tau_{\rm{b}}^2 }\!\!]
\end{equation}
with $\tau_{\rm{b}}$ being the bulk relaxation time.
The transition probability between quasiparticles of momentum $\mathbf{p}= (\mathbf q,p_z )$ and $\bf{p'}$ is given by
\begin{equation}
\label{W}
W(\textbf{p},\textbf{p}') = \frac {\pi^4 \hbar^2}{{m^{*2}} L^6} \,\zeta(\textbf{q} - \textbf{q}') \, \, \Big(\frac {p_{\rm{z}} L}{\pi \hbar}\Big)^2\, \, \Big(\frac {p'_{\rm{z}} L}{\pi \hbar}\Big)^2\,\,\,,
\end{equation}
where $m^*$ is the quasiparticle effective mass and ${\bf{q}}$ is the component of the quasiparticle momentum in the film plane and $p_z  = \pi \hbar j/L$.

We calculate the momentum transport relaxation time $\tau _\mathrm{osc}$ for the $^3$He film assuming a single rough surface and that the upper surface of the film is flat. In terms of the measured surface roughness spectrum $\zeta(\mathbf{k})$, the effective relaxation time for quasiparticles when coupled to a torsional oscillator is determined numerically from
\begin{equation}
\label{tau_osc_final}
\frac{1}{\tau_{\mathrm{osc}}} =\frac{1}{\tau_\mathrm{b}}\frac{k_{\rm{F}}} {L}\sqrt{k_{\rm{F}}\lambda_{\rm{in}}} \!\!\int\!\!\!\!\!\!\! \int_{S'}\!\zeta\!\!\left(k_x,k_y\right) \Gamma\!\left(\frac{k_x}{k_\mathrm F},\frac{k_y}{k_\mathrm F}\right)\mathrm dk_x\,\mathrm dk_y
\end{equation}
where $\Gamma(u,v)$ is the weight function for the k-space integral. The range of quasiparticle wave-vectors over which the integral is specified ($S'$) manifests the effect of confinement of the film in the $z$ direction: the transformation from a three-dimensional integral in Eq.~\eqref{MeyerovichFormulaTauEff} to a two-dimensional integral in Eq.~\eqref{tau_osc_final}.

    \begin{figure}[htp]
    \includegraphics[width=8.2cm]{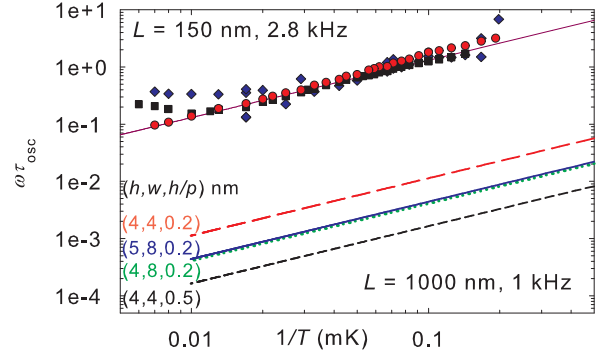}
    \caption{(color online)Top: The calculated and measured values of $\omega \tau_{\rm{osc}}$ for the undecorated surface showing good agreement for the momentum relaxation time extracted from the mesoscale surface profile, with film thickness 150 nm. Bottom: predicted values for various surface roughness profiles for a highly confined structure such as that encountered in \cite{dimov} but with engineered mesoscale roughness.  }
    \label{figure3}
    \end{figure}
The calculated $1/\tau _{{\rm{osc}}}$ shows a linear temperature dependence in good quantitative agreement with the experimental data, as shown in Fig.~\ref{figure3}. Thus these experiments confirm the essential validity of the theory \cite{Tesanovic86,TrivediAshcroft88,MeyerovichStepaniants} in which the independently experimentally measurable surface profile fully determines the effective disorder potential within the mesoscopic film, and thus allows a first principles calculation of the film transport (in the limit that elastic scattering processes within the film are negligible). Physically the observed behavior results from the interplay of weak elastic surface scattering and inelastic bulk scattering. The momentum transfer associated with elastic surface scattering (in the limit $1/\tau _\mathrm b  \to 0$) is determined by $\zeta (\mathbf q - \mathbf q')$. In the case of Gaussian correlations considered previously \cite{Meyerovich01}, this has a typical scale of order $\hbar/R$, where $R$ is the surface roughness correlation length of the atomically rough wall. Thus for surfaces for which the correlation length satisfies $p_\mathrm F R/\hbar \gg  1$,  $\Delta j \ll j$, wall scattering is rather ineffective at transverse momentum transfer. At finite temperature, the finite quasiparticle lifetime significantly increases the available phase space for surface scattering, and is responsible for the characteristic linear temperature dependence of the surface scattering term in the `interference' regime. This regime requires~\cite{Meyerovich01} that the bulk relaxation rate exceeds the wall relaxation rate $(\hbar/\tau _\mathrm b )/E_\mathrm F \gg (\hbar/R)^2 /p_\mathrm F^2$, or equivalently $\lambda_\mathrm{in}\ll k_\mathrm F^{ - 1} (k_\mathrm F R)^2$; in helium films this condition is satisfied at all temperature above 1~mK for our polished silver surfaces.

Experiments on helium films with a free surface suffer from a number of disadvantages, particularly a lack of precise control of the film thickness. Furthermore the upper free surface of the film is in practice significantly distorted due to van der Waals interactions with the substrate. The film morphology can be modelled so that its roughness power spectrum and the cross correlations between surface roughness of upper and lower surface may be included in a full calculation \cite{BowleyBenedict06}. However recent advances in the fabrication of nanofluidic cavities allow the fabrication of films of extremely well characterized thickness and fully characterizable surfaces~\cite{dimov}. Chemomechanical polishing of silicon can give correlation lengths in excess of one micron, and rms roughness of order 0.1nm, significantly better than achievable with our polished silver surfaces \cite{Teichert}. One of these highly smooth surfaces can be textured by lithographic techniques to engineer a required surface roughness. Our implementation of the theory applies to surfaces of arbitrary surface height profiles, satisfying $\xi (x,y)\ll L$. We are currently using this formulation of the theory in the design of cavities to probe the influence of nano-scale confinement on the superfluidity density of $^3$He.

In order to exhibit the generic dependence of the oscillator relaxation rate on the properties of the surface we write the height-height autocorrelation function for a statistically-isotropic random surface of correlation length $R$ as $\zeta(x,y)=l^2 f(x/R,y/R) $ where $f$ is a dimensionless function ($f(0,0)=1 $) describing the nature of the surface morphology; $l$  characterizes the height scale and $R$ the (in-plane) length scale. Correspondingly, the power spectrum may be written as $ \zeta(k_x,k_y)=l^2R^2g(k_xR,k_yR)$
where the dimensionless $g$ is the Fourier transform of $f$.

Then the oscillator relaxation rate, Eq~\eqref{tau_osc_final} can be cast in the general form
\begin{equation}\label{dimless-tau}
\frac{1}{\tau_\mathrm{osc}}=\frac{1}{\tau_\mathrm{b}}\sqrt{k_\mathrm F \lambda_\mathrm{in}}\left(\frac{k_\mathrm F l^2 }{L}\right) F(k_\mathrm F R)
\end{equation}
where $F(k_\mathrm F R) $ is a dimensionless morphology integral (for momentum relaxation).

Let us consider a model engineered sample surface composed of a lattice of narrow posts of rectangular cross-section, of height $h$, width $w$ and spacing/periodicity $p$. The surface roughness power spectrum of such a surface is shown in Fig.~\ref{figure2}d (dashed line). The mean-square height variation for such a surface is $l^2=h^2 w^2/p^2 $ and the correlation length is $R\sim w $. The calculated relaxation rate for $^3$He films within a cavity with this designed roughness is shown in Fig.~\ref{figure3}. The relaxation time is found to depend on $h^2/p^2$ for a fixed width $w$, as follows from Eq.\eqref{dimless-tau} in the limit $ w^2/p^2\ll 1$.

The key control parameters influencing mass transport of the film are the rms roughness of the surface profile $l$, the correlation length $R $ of the height variations and the nature of the surface roughness. The relaxation rate is simply proportional to $l^{2}$. The dependence on $R$ is through the dimensionless variable $k_\mathrm F R$, with the precise functional form determined by the weighting of different regions of $k-$space in the morphology integral $F(k_\mathrm F R) $; this, in turn, is determined   by the detailed surface roughness power spectrum. We also note that the particular form for the morphology integral will differ according to the transport response (e.g. mass transport, thermal conductivity) under consideration. For the case of a Gaussian spectrum the momentum relaxation rate has a simple form, proportional to $(l/R)^{2}$.

In conclusion, we have shown that $^3$He films provide a model mesocopic system with disorder potential fully determined by surface roughness. Transport in these relatively thick films already manifests quantum size effects via this disorder potential. The anomalous temperature dependence of the transport time is theoretically accounted for and can be calculated from the measured surface roughness power spectrum and known inelastic relaxation time. The experimental demonstration that the relaxation rate depends not only on vertical roughness scale but also on lateral correlations has potential implications for the engineering of high quality metallic nanoscale interconnects in electronic devices.

We thank Roger Bowley for useful discussions and acknowledge support from the EPSRC grants GR/S20677 and EP/E054129, European Microkelvin Consortium (FP7 grant agreement no:228464), the National Science Foundation DMR-0806629 and the Leverhulme Trust.


\begin{thebibliography}{999}

\bibitem{Josell}
D. Josell, S.H. Brongersma, and Z. Tokei, Annu. Rev. Mater. Res., \textbf {39}, 231 (2009).

\bibitem{Fuchs}
K. Fuchs, Proc. Camb. Philos. Soc., \textbf{34}, 100 (1938)

\bibitem{Sondheimer}
E.H. Sondheimer, Adv. Phys., \textbf{1}, 1 (1952).

\bibitem{Tesanovic86}
Z. Tesanovic, M. V. Jaric and S. Maekawa, Phys. Rev. Lett. \textbf{57}, 2760(1986).

\bibitem{TrivediAshcroft88}
N. Trivedi and N. W. Ashcroft, Phys. Rev. B \textbf{38}, 12298 (1988).

\bibitem{MeyerovichStepaniants}
A.E. Meyerovich and A. Stepaniants, Phys. Rev. B \textbf{58}, 13242 (1998). Phys. Rev. B \textbf{60}, 9129 (1999). J. Phys. Cond. Mat \textbf{12}, 5575 (2000).

\bibitem{MeyerovichPonomarev}
A.E. Meyerovich and I.V. Ponomarev, Phys. Rev. B \textbf{65}, 155413 (2002)

\bibitem{MeyerovichChatterjee}
S. Chatterjee and A.E. Meyerovich, Phys. Rev. B \textbf{81}, 245409 (2010)

\bibitem{Munoz}
R.C. Munoz, C. Arena, G. Kremer and L Moraga, J. Phys. Cond. Mat. \textbf{15}, L177 (2003)

\bibitem{Feldman}
B. Feldman, R. Deng and S.C. Dunham, J. App. Phys. \textbf{103}, 113715 (2008).

\bibitem{Sun}
T. Sun et al., Phys. Rev. B \textbf{81}, 155454 (2010)


\bibitem{Meyerovich01}
A. E. Meyerovich,  JLTP \textbf{124} (3/4), 461(2001).

\bibitem{Casey1}
A. Casey, J. Parpia, R. Schanen, B. Cowan and J. Saunders, Phys. Rev. Lett., \textbf {92} 255301 (2004).

\bibitem{dimov}
S. Dimov, et al. Rev. Sci Inst. \textbf{81} 01390 (2010).

\bibitem{BowleyBenedict06}
R. M. Bowley and Keith A. Benedict, JLTP \textbf{142}, 701 (2006).

\bibitem{Teichert}
C. Teichert et al., App. Phys. Lett. \textbf{66}, 2346 (1995)

\bibitem{natural}
 H. W. Deckman and J. H. Dunsmuir, J. Vac. Sci. Technol. B \textbf{1}, 1109 (1983).


     \end{thebibliography}
\end{document}